\documentstyle[12pt,cite,times,fleqn,twoside,epsfig,rotating]{article}
\begin{document}
\newpage
\vspace{2cm}
\hspace{12cm} {\bf DESY 97-170}
\vspace{2cm}
\begin{center}
{\LARGE {\bf
 Study of $e^+ e^- \rightarrow WW \rightarrow l \nu q \bar{q}$ at 500 GeV}} \\
\vspace{2cm}
\it{Contribution to the Proceedings of the "Joint ECFA / DESY Study: Physics and Detectors for a Linear
Collider", February to November 1996, DESY 97-123E, Hamburg 1997}
\end{center}
\vspace{1cm}
\samepage
\nopagebreak
\newcount\tutecount  \tutecount=0
\def\tutenum#1{\global\advance\tutecount by 1 \xdef#1{\the\tutecount}}
\def\tute#1{$^{#1}$}
\tutenum\athens            
\tutenum\barcel            
\tutenum\zeuthen           
{
\parskip=0pt
\noindent
\ifx\selectfont\undefined
 \baselineskip=11.2pt
 \baselineskip\baselinestretch\baselineskip
 \normalbaselineskip\baselineskip
 \ixpt
\else
\selectfont
\fi
\medskip
\tolerance=10000
\hbadness=5000
\raggedright
\hsize=162truemm\hoffset=0mm
\def\r{\rlap,}
\noindent
{\bf
George Daskalakis\r\tute{\athens}\
Aristoteles Kyriakis\r\tute{\athens}\
Christos Markou\r\tute{\athens}\
Hannelies Nowak\r\tute{\zeuthen}\
Imma Riu\r\tute{\barcel}\
Errietta Simopoulou\r\tute{\athens}\
Revaz Shanidze\r\tute{\zeuthen}\ }
\rule{\textwidth}{0.4pt}
\medskip
\vspace{1.2cm}
\begin{list}{A}{\itemsep=0pt plus 0pt minus 0pt\parsep=0pt plus 0pt minus 0pt
                \topsep=0pt plus 0pt minus 0pt}
\item[\athens]
  National Center for Scientific Reseach "Democritos", Aghia Paraskevi,
  Attikis, Greece
\item[\barcel]  Institut de Fisica d'Altes Energies,
     Universitat Autonoma de Barcelona,
      E-08193 Bellaterra (Barcelona), Spain
\item[\zeuthen] DESY-Institut f\"ur Hochenergiephysik, D-15738 Zeuthen,
     Germany
\end{list} }
\samepage
\vspace{4cm}
\nopagebreak
\samepage
\begin{abstract}
W-pair production in $e^+e^-$ annihilation at 500 GeV is studied using
different Monte Carlo generators and the proposed detector for the
Linear Collider
in its TESLA version. Semileptonic W decays are used to give hints for
the detector optimization and to determine the expected accuracy of the
W mass determination and the measurement of the anomalous triple
gauge boson couplings. With an expected luminosity of  50~$fb^{-1}$ a W mass
measurement with an accuracy of 15 MeV should be possible. The triple
gauge boson couplings can be determined with an accuracy in the order of 10~$^{-3}$. \\
\end{abstract}
%
\newpage
\section{Cross-Section of WW Pair Production and Event Rates}

\begin{figure}[h]
\vspace{-2cm}
\mbox{\hspace{-2.5cm}{\epsfig{figure=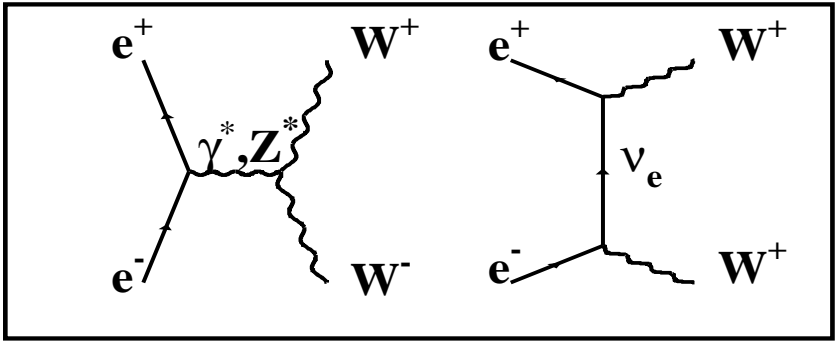,width=20cm,height=25cm}}}
\vspace{-18.0cm}
\caption{ \label{fig:wwgroup:diag}
Lowest order Feynman diagrams for W-pair production in 
$e^+e^-$ annihilation }
\end{figure}

   The W-pair production in $e^+e^-$ annihilation is described in the
 Standard Model by three lowest
 order diagrams: two $s$ channel diagrams containing the
 triple gauge boson couplings and one $t$ channel diagram with $\nu_e $
 exchange, that contributes to  left-handed electrons only.
These diagrams are presented in figure \ref{fig:wwgroup:diag}. \\

 In the context of this study we have compared
 the total cross-sections obtained from different
Monte Carlo generators in the Born approximation
 \cite{wwgroup:comphep,wwgroup:erato,wwgroup:wopper,wwgroup:koralw,wwgroup:exali}.
 The Standard Model  parameters for the cross-section calculations
as used
  in the different Monte Carlo generators are summarized in table
\ref{wwgroup:tab:param}.\\
\nopagebreak
\vspace{-1cm}
\begin{table}[h]
\begin{center}
\caption {
   Input parameters and obtained cross-section at 500 GeV for
      W-pair production in $e^+ e^-$ annihilation}\label{wwgroup:tab:param}

\vspace{0.8cm}
\begin{tabular}{|c||c|c|c|c|c|} \hline \hline

 Parameters & CompHEP &  WOPPER & KORALW & ERATO&  EXCALIBUR  \\ \hline \hline
  Z mass (GeV) &  91.187 &  91.181  & 91.181   & 91.19  & 91.19    \\
  Z width(GeV) &   2.502 &   2.501  &  2.501   & 2.497  &  2.497   \\
  W mass (GeV) &   80.29 &   80.44  &  80.44   &  80.23 & 80.23    \\
  W width(GeV) & 2.094   &   2.10  &   2.10    &  2.03  & 2.03     \\
  $sin^2\theta_W$ & 0.22468 & 0.22995 & 0.22995 &  0.23103  & 0.23103 \\
   1/$\alpha$  &  128.0 & 128.00 & 128.00 & 128.07 & 128.07 \\  \hline \hline
  cross-section &  7.556 & 7.530 & 7.534  &   7.2  &   7.356 \\ \hline \hline
          
\end{tabular}
\end{center} 
\end{table}

\begin{figure}[ht]
\vspace{-1cm}
\mbox{\epsfig{figure=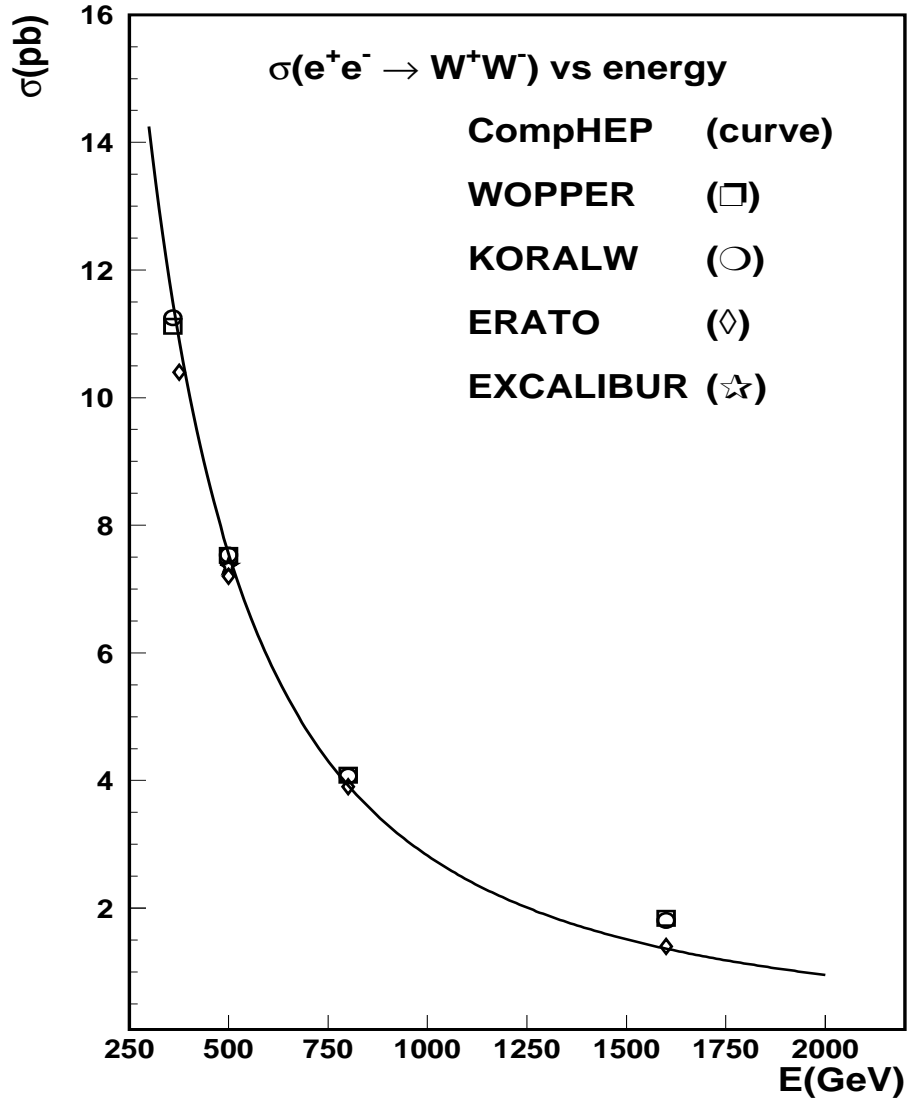,width=14cm,height=18cm}}
\vspace{-0.5cm}
\caption{\label{fig:wwgroup:totx}
Total cross-section of WW pair production }
\end{figure}

 The results of the different generators
at $\sqrt{s}=$ 360, 375, 500, 800 and 1600 GeV are shown in figure \ref{fig:wwgroup:totx}.
The full line represents the results from the CompHEP program.
All cross-sections were calculated for CC03 diagrams without beamstrahlung and initial
state radiation (ISR). No cuts were applied to the data.
All cross-sections agree within a few percent. \\
\pagebreak

The cross-section including
 initial state radiation and beamstrahlung effects
 is in the order of 8~pb at 500 GeV. Assuming
  2.5 years running period at this energy, 50~$fb^{-1}$ of luminosity  will be collected.
This corresponds to 400000 WW pairs.
  To determine triple gauge boson couplings from WW events in
 $e^+e^-$ annihilation, the charge of the W has to be known. This is
 defined through the charge of the lepton.
Taking into account the branching ratios of $W^- \rightarrow e^- \nu $ and
$W^- \rightarrow \mu^- \nu$, about 30000 events of each decay are expected. \\

\newpage
\section{Generation of W Pairs and Background Reactions}

\subsection{Generation and Selection of W Pairs }

\begin{figure}[h]
\vspace{-0.6cm}
\mbox{\epsfig{figure=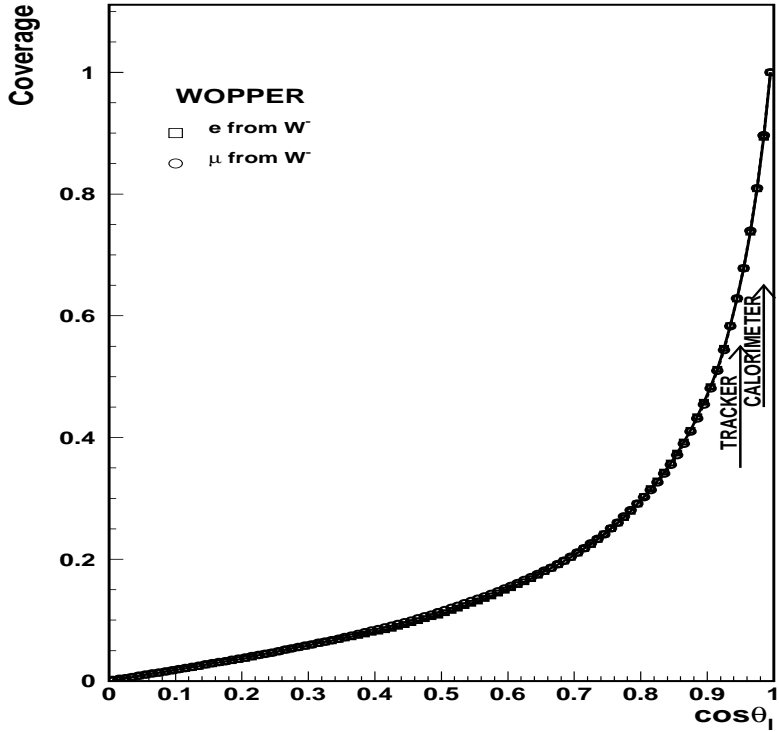,width=12cm,height=12.cm}}
 \caption{ \label{fig:wwgroup:cover}
Detector coverage as a function of $cos\theta_{l}$. The sensitivity
limits for the inner tracker and the calorimeter are indicated by arrows.
}
\end{figure}

\begin{figure}[h]
\vspace{-1.0cm}
 \mbox{\hspace{1.5cm}{\epsfig{figure=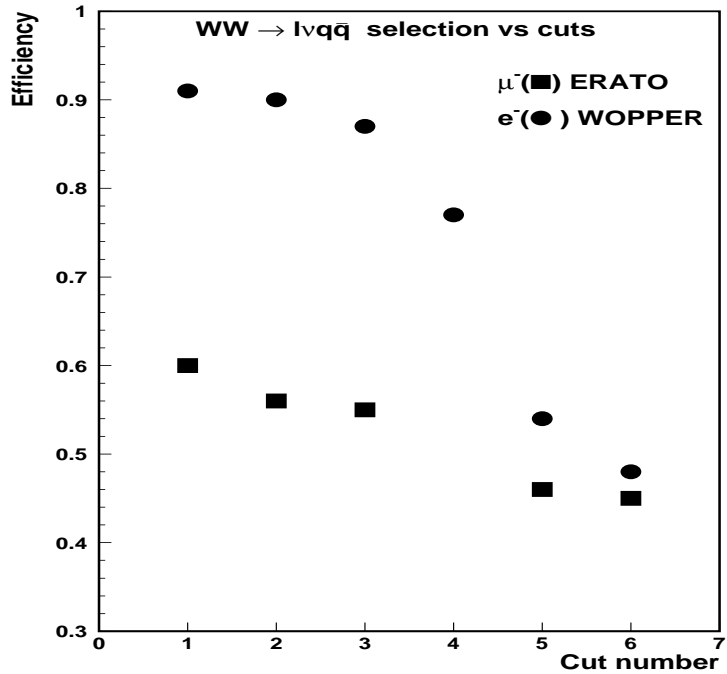,width=11cm,height=11.cm}}}
\vspace{-0.6cm}
\caption{\label{fig:wwgroup:effs}
Selection efficiencies as a function of the selection cuts}
\end{figure}

 WW pairs were generated with two different generators ERATO \cite{wwgroup:erato}
and WOPPER \cite{wwgroup:wopper}. The triple gauge boson
couplings were set to the Standard Model
values. Initial state radiation according to \cite{wwgroup:isr}
and beamstrahlung using CIRCE
 \cite{wwgroup:circe} for the
TESLA option were taken into account. \\
  The most relevant variable for the selection efficiency of
$e\nu q\bar{q}$ and $\mu \nu q\bar{q}$ final states is $cos\theta_{l}$
where $\theta_{l}$ is the polar angle of the produced lepton.
The dependence of the detector coverage as a function of $cos\theta_{l}$
at generator level is illustrated in fig. \ref{fig:wwgroup:cover}.
\nopagebreak
Due to the extremely forward production of the W bosons
the maximum efficiency as given by the detector geometry is about 85 \%
for the calorimeter and  60\% for the geometry of the inner tracker. \\
 34000 WW pairs of the final state $ \mu{^\pm} \nu $ $ q \bar{q}$
 were generated with ERATO using CC03 diagrams and
30000 WW pairs decaying  into $ e^- \nu q \bar{q}$ as well as
and 30000 WW pairs decaying into $ \mu^- \nu q \bar{q}$  with
WOPPER including CC03 and CC11 diagrams. \\
Muons were analyzed in the inner tracker,
electrons in the inner tracker , the calorimeter and in the luminosity detector
if they hit it. \\
The isolation criteria were as follows:  For muons, we
required the angle between 
\pagebreak
the muon and the closest jet to be larger than
$30^o$. For electrons, we required less than 10 GeV of energy in a cone
of $23^o$.
The final cuts for the
selection of WW final states with electrons and muons are presented
in table \ref{wwgroup:tab:selec} and figure \ref{fig:wwgroup:effs}.
The selection efficiency for the electron
 set of cuts is 0.480 $\pm$ 0.003 while for the 
muon set it is 0.450 $\pm$ 0.003. \\
\nopagebreak
\begin{table}[h]
\begin{center}
\vspace{-0.8cm}
\caption{  Selection cuts as used in the WW analysis }
\label{wwgroup:tab:selec}
\vspace{0.5cm}
\begin{tabular}{|c||c|c|c||c|c|c|}\hline \hline
 variable & cut num.  &value (e sel.)& eff. &cut num.  & value ($\mu$ sel)& eff. \\ \hline  \hline
 cos$(\theta_{l}$)& 1 &$\le 0.9848$+lumi & 0.91 & 1 & $\le 0.95  $& 0.60 \\ \hline
$p_l $ & 2 &$\ge 20$ GeV  & 0.90 & 2 & $\ge 20$ GeV  & 0.56   \\ \hline
$\alpha_{iso}$ & 3 & $\ge  23^o$ & 0.87  & 3 &$\ge 30^o $& 0.55   \\ \hline
$E_{had}$& 4 &$\ge 0.2 \ast  \sqrt{s}$  & 0.77    &  &  &    \\ \hline
M(lmiss)  & 5  & 50 - 110 GeV& 0.54    & 5  & 60 - 100 GeV& 0.46 \\ \hline
M(jj)     & 6  & 60 - 100 GeV& 0.48    & 6  & 60 - 100 GeV& 0.45 \\ \hline  \hline
\end{tabular}
\end{center}
\end{table}
\vspace*{-0.4cm}
 \nopagebreak

\newpage

\subsection{Background simulation}

We have considered two different sources of background:
background  from the other decay channels of the  WW pairs  and
background  from
processes  like $ee \rightarrow ZZ, t\bar{t} , q \bar{q} $.
   All these processes were generated at 500 GeV using PYTHIA \cite{wwgroup:pythia}
 , including ISR and beamstrahlung.
All background contributions taken into account are summarized
in table \ref{wwgroup:tab:backg}.
   The total amount of these backgrounds is 2.24\%. \\
\nopagebreak
\begin{table}[h]
\begin{center} 
\vspace{-1.2cm}
\caption{
 Background contributions  to $e^+e^- \rightarrow W^+W^-  \rightarrow l \nu q
 \bar{q}$ events from the different reactions, corresponding to the
 integrated luminosity of 50 $fb^{-1}$ }\label{wwgroup:tab:backg}
\vspace{0.5cm}
\begin{tabular}{||r||c|c|c||} \hline \hline

  Reaction~~~~  & cross-section & events  &  rate  \\
 $e^+e^- \rightarrow$~~~~ & pb  &  after cuts &  $\%$  \\  \hline \hline
 $WW \rightarrow q \bar{q} q\bar{q}$  &  3.6 &  -   & -   \\ 
 $l \nu l \nu $  &  0.8  &  - & -   \\  
  $\tau^- \nu q \bar{q}$  &  0.6  &  260 & 1.8  \\ \hline 
 $ q \bar{q} $~~~~~~~~    &  12.0 &  3 & 0.02    \\  
 $ t \bar{t}$~~~~~~~~     &  0.6  &  3 & ~~0.02~~    \\ 
 $ZZ$~~~~~~~~             &  0.4  &  50 & 0.4    \\  \hline
  all                     &       &     & 2.24   \\ \hline  \hline
\end{tabular}
\end{center}
\end{table}

\vspace{-0.6cm}
\subsection{Detector Model}

 The detector model that we have used in our analyses was
 developed at DESY-IfH Zeuthen  and is described
 elsewhere \cite{wwgroup:fastmc}.
 The following parametizations for momentum and energy resolution
 were used in the simulation program:
 \begin{eqnarray}
 {\delta{p_t}\over{{p_t}^2}} & = & {a \times10^{-4}{(\frac{GeV}{c})}^{-1}} \\
 {\delta E \over E} &  = & {{a\over\sqrt{E}} \oplus b} .
 \end{eqnarray}
  In table \ref{wwgroup:tab:detec} are summarized the main parameters
used in this fast simulation program.  \\
 All generated events had to pass this program before any analysis.
\nopagebreak
\begin{table}[h]
\begin{center}
\vspace{-0.9cm}
\caption{Detector parameters used in simulation }
\label{wwgroup:tab:detec}
\vspace{0.3cm}
\begin{tabular}{|l||c|c|c|c|c|c|} \hline \hline
& Coverage & Threshold & \multicolumn{2}{c|}{Granularity}
& \multicolumn{2}{c|}{Resolution} \\ \cline{2-7}
\raisebox{1.5ex}[0cm][0cm]{Detector} & $|cos \theta| <$ & $E>$(GeV)&
$\Delta \theta^o $ & $\Delta \phi^o$ & $a $  & $ b $  \\  \hline \hline
inner tracker   &  0.95   & 0.2 & -   &  -   &  1.5 &  -  \\  \hline
ECAL            &  0.985  & 0.1 & 0.7 &  0.7 &  0.10 & 0.01  \\  \hline
HCAL            &  0.985  & 0.3 & 2   &  2   &  0.50 & 0.04  \\  \hline
$\mu$ detector  &  0.95   & 0.2 &     &      &  1.5  &     \\  \hline
LUMI detector  &   0.9976 & 30  & 1   &  1   &  0.10 & 0.01  \\  \hline
\hline
\end{tabular}
\end{center}

\end{table}


\newpage
\section{ W Resolution} \vspace{0.3cm}
 
\begin{figure}[h]
\vspace{-1cm}
 \mbox{\epsfig{figure=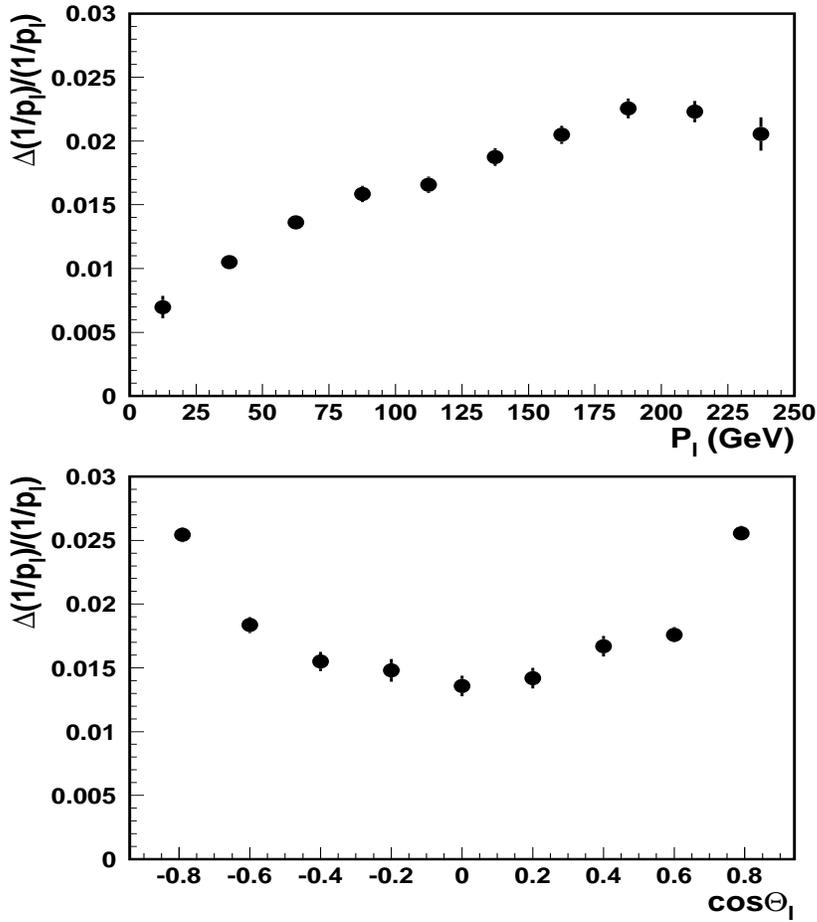,width=12cm,height=15.cm}}
\vspace{-0.8cm}
\caption{ \label{fig:wwgroup:moml}
 (a) lepton momentum resolution as a function of (a) the momentum
 and (b) cos$\theta_l$ }
\end{figure}

 The resolution of the lepton momentum as a function of the momentum 
itself as well as a function of cos$\theta_l$ is shown in fig.
\ref{fig:wwgroup:moml}. \\
\nopagebreak
A cut of 20 GeV was introduced to exclude leptons
from beauty and charm  decays.
As can be seen from this figure, the momentum resolution for leptons
with a momentum larger than 20 GeV is in the order of 2 \%.
This is true for the whole  inner tracker ($|cos \theta_l| \le$ 0.83 ).
This means that we have no charge mismatch for these events. \\

\begin{figure}[h]
\vspace{-1.5cm}
 \mbox{\epsfig{figure=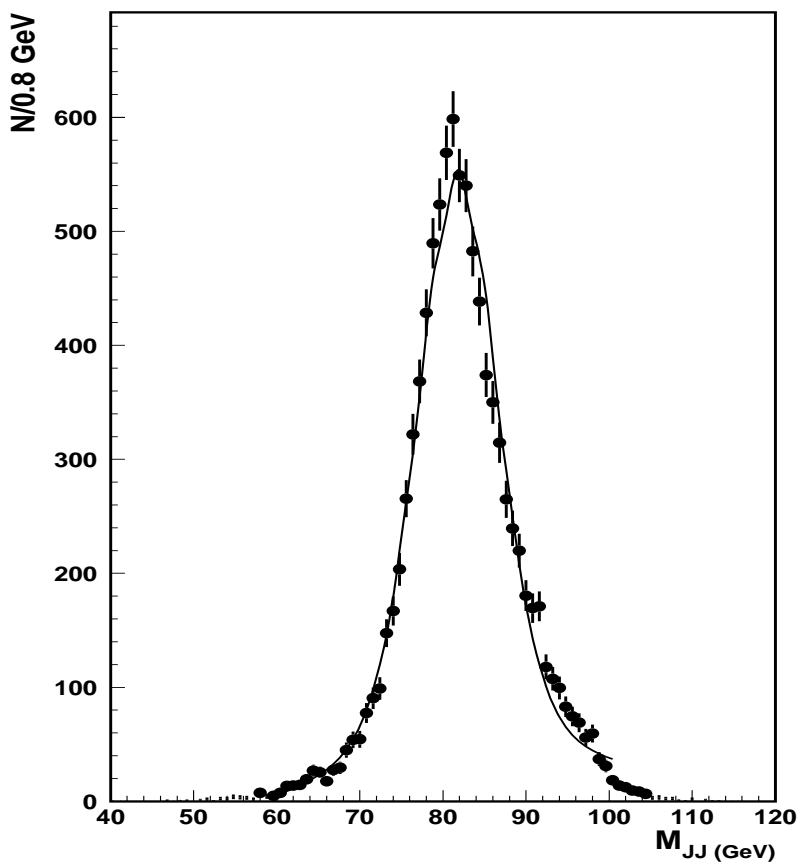,width=12cm,height=14.cm}}
\caption{ \label{fig:wwgroup:bwmf}
 W mass spectrum from invariant jet-jet mass for the muon final state }
\end{figure}

\begin{figure}[ht]
\vspace{-1cm}
 \mbox{\epsfig{figure=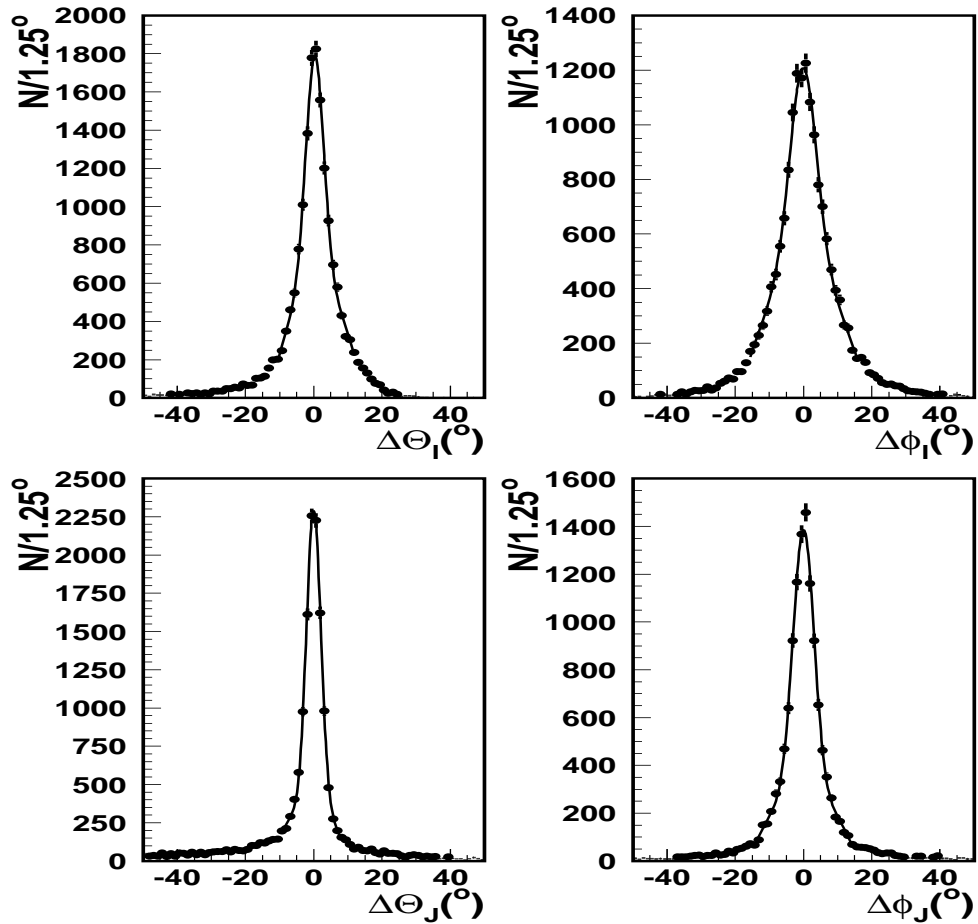,width=15cm,height=15.cm}}
\caption{ \label{fig:wwgroup:ares}
 Angular resolutions of the detector for lepton and jet angles }
\end{figure}

Events selected with these cuts were  forced to have two jets.
 The Durham algorithm \cite{wwgroup:durham}
was used for the jet reconstruction.
 Then a kinematical fit was performed requiring four-momentum
conservation and equal masses of the two W bosons.
  From the selected events two invariant masses can be
 calculated: the mass of the hadronic system is the
 two jet invariant mass $M_{jj}$ and $M_{l \nu}$ is the invariant mass of
 the isolated lepton and the missing momentum.
The relevant variable for the measurement of the W mass is $M_{jj}$
because it is better determined and its errors are smaller.
 \begin{figure}[h]
 \mbox{\epsfig{figure=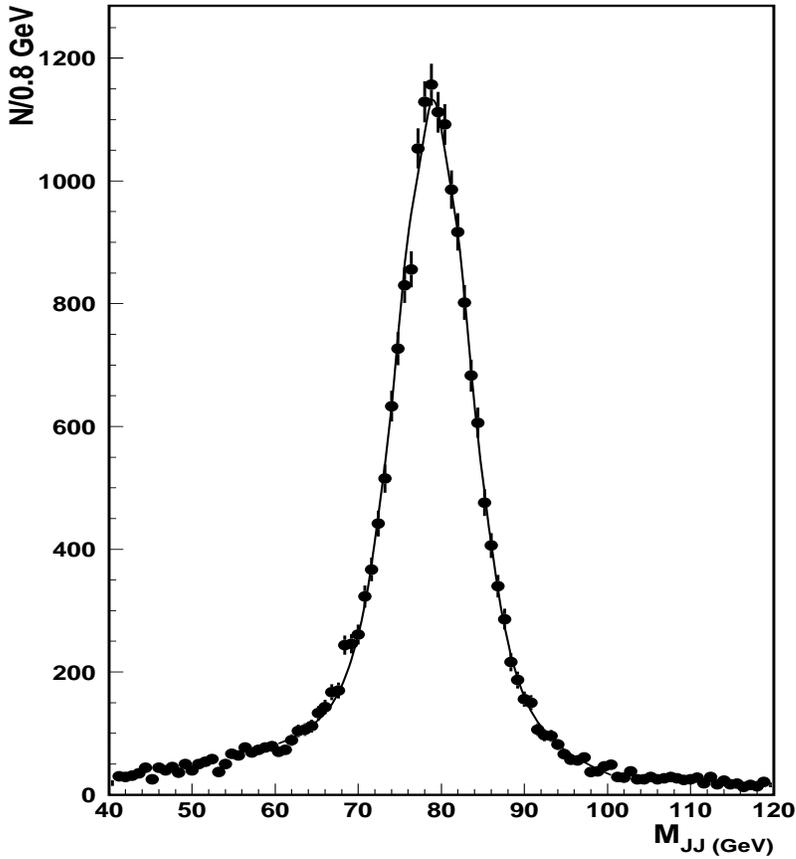,width=12cm,height=14.cm}}
\caption{ \label{fig:wwgroup:bwef}
 W mass spectrum from invariant jet-jet mass for the electron final state }
\end{figure}
    Fig. \ref{fig:wwgroup:bwef} shows the mass plots for the invariant jet-jet
mass  for the electron  and fig. \ref{fig:wwgroup:bwmf} for the
muon  final state.  \\
In the fitting procedure
a Breit-Wigner function convoluted with a Gaussian was used as
can be seen in the following formula:

\begin{eqnarray}
  M(x,m_{W},\Gamma_{W},\sigma) & = &
     {\int_{y_{1}}^{y_{2}}}BW(y;m_{W},\Gamma_{W})G(y-x,\sigma)dy. \label{eq:bwg}
\end{eqnarray}
The width of the Breit-Wigner was fixed to the width of the W as used
in the Monte Carlo generator. The $\sigma$ of the Gaussian represents the
mass resolution of the apparatus. We obtained mass resolutions of 3.6
and 3.9 GeV for electrons and muon final states, respectively.
 The limits for the integration are
chosen to be four times the experimental resolution.
The mean value itself is shifted to lower values in the case of the hadronic
mass. This is due to the wrong assignment of particles inside a jet,
the loss of some particles outside of the acceptance
of the detector, detector thresholds and the ISR photon. These effects
can be corrected statistically.
The error of the mean mass value represents a measure of the quality of
the mass measurement.
The minimal error of $M_W$ is given by the following relation:
\begin{eqnarray}
    \Delta M_W \sim   \frac{\Gamma_W}{\sqrt{N}}.   \label{eq:dm}
\end{eqnarray}
 This is true for an ideal detector and an efficiency of 100\%.
 Replacing ${\Gamma_W}$ by the mass resolution (4.0 GeV)
and taking
into account the detection efficiency of about 35 \%
leads to
an expected error of the W mass of about 18 MeV using the full available
statistics of semileptonic final states ( $e^{\pm}$,$\mu^{\pm}$ and 
$\tau^{\pm}$). \\
\newpage
A further increase of
statistics will not decrease this value significantly because of the
dominance of the detector systematics. Better understanding of this
systematics and the fitting procedure can improve this error slightly.\\
The resolution of
$\theta_W$ , $\cos\theta_l$ and $\phi_l$ as well as $\cos\theta_q$
and $\phi_q$ is relevant for the determination of the anomalous couplings.
The last four variables are defined in the W rest frame. 
The corresponding values are presented in table \ref{wwgroup:tab:angle}.
\begin{table}[h]
\begin{center}
\vspace{-6mm}
\caption {
   Resolutions for W production angle and
      its decay products angles  
        in the W rest frame }\label{wwgroup:tab:angle}
\vspace{0.1cm}
\begin{tabular}{||c||c|c||}  \hline \hline
resolution of       & $\mu$ sel.&  e sel.   \\ \hline \hline
$cos \Theta_W $     &  0.0083  &   0.0082  \\ \hline
 $ \Theta_W $    &  1.20  &   1.19  \\  \hline
$cos \Theta_q $     &  0.026   &   0.027   \\ \hline
 $ \Theta_q $   &  2.5   &    2.6  \\    \hline
 $\phi_q$     &   3.0    &   2.8     \\ \hline
$cos \Theta_l $     &  0.050   &   0.067   \\ \hline
 $ \Theta_l$    &  2.7   &   2.6   \\   \hline
  $\phi_l$    &   4.0    &   3.9     \\ \hline \hline
\end{tabular}
\end{center}     
\end{table}
\samepage
\vspace{-0.5cm}
\nopagebreak
 \begin{figure}[h]
\vspace{-1cm}
 \mbox{\hspace{1.5cm}{\epsfig{figure=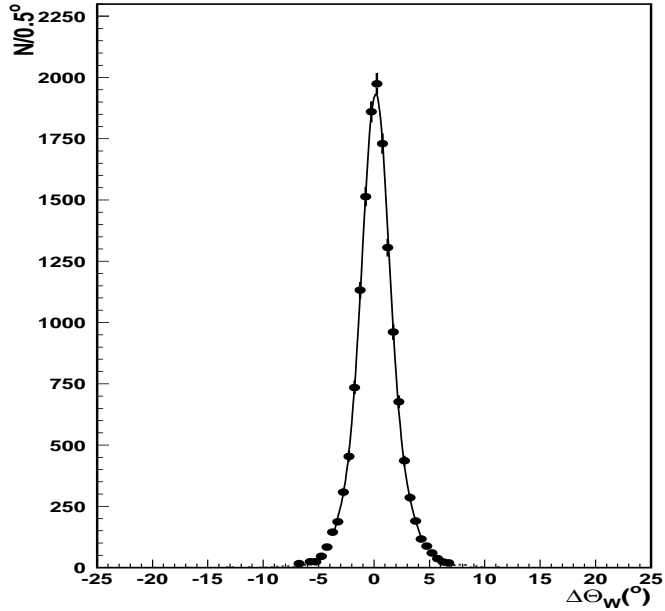,width=10cm,height=10.cm}}}
\caption{ \label{fig:wwgroup:angleW}
 Angular resolutions for $\theta_W$ as reconstructed from the two jets}
\end{figure}
\newpage
 The angular resolutions are all in the order of few degrees.
The corresponding angular distributions are presented in figure
\ref{fig:wwgroup:ares}. Each distribution was fitted simultaneously with
 a gaussian  to
describe the W and  a second order polynomial to describe the background.
\nopagebreak
\newpage

\section{ Limits on the Anomalous Couplings}

   A maximum likelihood fit has been used to evaluate the sensitivity
of the anomalous coupling measurements at 500 GeV.
Monte Carlo events were produced using the ERATO generator and passed
through the detector simulation and the analysis program. Only CC3 diagrams
with beamstrahlung, initial state radiation and Coulomb corrections
were included.
The matrix element $M_i$  for each event passing the
selection criteria mentioned above was calculated with the
ERATO generator using the 4-momenta of the lepton, neutrino and two jets after the
constraint fit.
 The minimization function is given
by:
\begin{eqnarray}
    -\log \prod_{i=1}^{N_{events}}\frac{|M_i|^{2}}{\sigma_{tot}}. \label{eq:log}
\end{eqnarray}

Quarks are not identified so a folding has to be done taking into account
the proper symmetrization of the matrix element squared.
The measured total
cross-section is denoted by $\sigma_{tot}$. 
The $\sigma_{tot}$ was calculated using the final selection cuts for the muon
decay channel.
It will be parametrized
according to ref. \cite{wwgroup:papa} as :
\begin{eqnarray}
              \sigma_{tot} & = &
  S_0 + S_1 \cdot \alpha + S_2  \cdot \alpha ^2. \label{eq:sig}
\end{eqnarray}
where $\alpha$ denotes generically any of the anomalous couplings investigated.
One parameter fits were performed for different
anomalous couplings. All of them are expected to be zero in the
Standard Model.
The results  are given in the table \ref{wwgroup:tab:actab}
following the notation of \cite{wwgroup:bile}.
\vspace{-0.8cm}
\begin{table}[h]
\begin{center}
\caption{
   Results  on accuracy of TGB couplings estimation from ERATO
    }\label{wwgroup:tab:actab}
\vspace{0.3cm}
\begin{tabular}{||c||c|c||}  \hline \hline
    Model    &  Coupling  &  Value  \\  \hline \hline      
 $\hat O_{B_{\phi}}$ & $ x_{\gamma}$ & (0.0 $\pm$ 2.8)$\times 10^{-3}$ \\
  $\hat O_{W_{\phi}}$  &  $ x_{\gamma} $ & (0.0 $\pm$ 1.4)$\times 10^{-3}$~~~~  \\
 $\hat O_{W}$  &  $\lambda_{\gamma}$ & (0.0 $\pm$  1.7) $\times 10^{-3}$ \\ \hline \hline
\end{tabular}
\end{center}     
\end{table}
\nopagebreak
The obtained values are in good agreement with the results of reference  \cite{wwgroup:pap}.\\
\nopagebreak
\section {Conclusions}
\nopagebreak
\begin{itemize}
\item The proposed detector is well suited for W physics.
\item Better equipped forward-backward regions will improve the
      WW results significantly.
\item A conservative estimate of the reachable error of the W mass
      measurement is 20 MeV.
\item A conservative estimate of the errors of the triple gauge
      boson couplings shows that they are in the order of $10^{-3}$.
\item This will improve over the results expected from LEP2 assuming an
      integrated luminosity of 500 $pb^{-1}$, by a factor of
      5 for the W mass and by an order of magnitude for the anomalous
      couplings.
\end{itemize}
\newpage

\end{document}